\documentclass[conference,a4paper]{IEEEtran}
\usepackage{float}
\usepackage{amsmath}
\usepackage{amsthm}
\usepackage{amssymb}	
\usepackage{graphicx}
\usepackage{subfigure}
\usepackage{enumitem}
\usepackage{algorithm}
\usepackage{stfloats}
\usepackage{cite}
\usepackage{cuted}
\usepackage{color}
\newcommand*{\J}{\jmath}%
\usepackage{amsmath,amssymb,mathtools,bm,etoolbox}
\newtheorem{my_theorem}{Theorem}

\newtheorem{my_lemma}{Lemma}

\newtheorem{my_proposition}{Proposition}

\usepackage{color}
\usepackage{mathtools}
\usepackage{suffix}
\usepackage{breqn}
\usepackage{hhline}
\usepackage{cite}
\DeclarePairedDelimiterXPP\Aver[1]{\mathbb{E}}{[}{]}{}{
	
	#1
}
\DeclarePairedDelimiterX\MeijerM[3]{\lparen}{\rparen}%
{\,#3\delimsize\vert\begin{matrix}#1 \\ #2\end{matrix}}
\newcommand\MeijerG[8][]{%
	G^{\,#2,#3}_{#4,#5}\MeijerM[#1]{#6}{#7}{#8}}
\newcommand{\diff}{\mathop{}\!d}

\WithSuffix\newcommand\MeijerG*[7]{%
	G^{\,#1,#2}_{#3,#4}\MeijerM*{#5}{#6}{#7}}
\makeatletter
\def\blfootnote{\xdef\@thefnmark{}\@footnotetext}
\makeatother
\usepackage{geometry}
\title{Statistical Results of Multivariate Fox-H Function for Exact Performance Analysis of RIS-Assisted Wireless Communication} 
\author{
	\IEEEauthorblockN{ Vinay Kumar Chapala and S.~M.~Zafaruddin}\\
	\IEEEauthorblockA{ Department of Electrical and Electronics Engineering, 
		BITS Pilani, Pilani Campus, Pilani-333031, Rajasthan, India
		\\ Email: \{p20200110, syed.zafaruddin\}@pilani.bits-pilani.ac.in}

}

\begin{document}
	\maketitle
	\begin{abstract}
Existing research provides statistical results on the sum of single-variate Fox-H functions to analyze the performance of diversity receivers and reconfigurable intelligent surfaces (RIS) based wireless systems. There is a research gap in exact performance analysis when more than a single-variate Fox-H function represents the statistical characterization of wireless systems. In this paper, we propose a novel approach to obtain the distribution of the sum of independent and non-identically distributed (i.ni.d) random variables characterized by the multivariate Fox-H function. Further, we develop a general framework for an exact analysis of the ergodic capacity when the multivariate Fox-H function characterizes the statistics of signal-to-noise ratio (SNR). We apply the derived results to conduct an exact performance analysis of outage probability and ergodic capacity, taking an example of  RIS-assisted communication over Rician fading channels with phase errors. We conduct computer simulations to validate the exact analysis and demonstrate performance of the RIS-assisted system under various practically relevant scenarios for a better performance assessment.
	\end{abstract}
%	\newpage
	\blfootnote{This work was supported in part by the Science and Engineering Research Board (SERB), Government of India, under MATRICS Grant MTR/2021/000890.}
	\section{Introduction}
Reconfigurable intelligent surfaces (RISs) have garnered significant interest from academia and industry due to their potential to control the wireless channel, which was previously considered uncontrollable \cite{Basar2019_access}. They are seen as strong contenders for beyond fifth-generation (B5G) and sixth-generation (6G) wireless networks. RISs consist of passive reflecting elements on a planar metasurface, capable of independently manipulating incident signals in real-time through phase shifts and amplitude attenuation. These surfaces offer improved spectral and energy efficiency, utilizing low-cost hardware and eliminating the need for self-interference cancellation or relaying techniques \cite{Yishi_2021, Qingqing2021}. 
In order to conduct statistical performance analysis of RIS-assisted wireless communications, it is crucial to ascertain the distribution for the sum of the product of two fading coefficients corresponding to the two links.

There has been extensive research on the distribution of the sum of non-identical fading coefficients for wireless communications over various channel conditions \cite{digital2005}. Conventionally, the sum of random variables is encountered in the analysis for diversity receivers such as equal-gain combining (EGC) and maximal-ratio combining (MRC).   Analyzing the performance of RIS-assisted systems is more challenging compared to diversity receivers due to the need to analyze the sum of double-fading channel coefficients. The authors in \cite{Abo2018}  presented  exact closed-form expressions for the probability density function (PDF) and cumulative distribution function (CDF) of the sum of an arbitrary number
of independent and non-identically distributed (i.ni.d) single-variate Fox-H function random variables in terms of the multivariate Fox’s H-function.
Motivated by this, researchers in \cite{trigui2020_fox, du2020_thz, chapala2021THz, chapala2021unified,chapala2022reconfigurable} employed the multivariate Fox-H function representation to develop exact analyses for various RIS-assisted wireless networks. In \cite{trigui2020_fox}, an exact performance analysis for RIS-assisted wireless transmissions over generalized Fox's H fading channels was presented. In \cite{du2020_thz, chapala2021THz}, RIS-aided THz communications were analyzed considering different fading channel models, antenna misalignment, and hardware impairments. In \cite{chapala2021unified}, a unified performance analysis for a free-space optical (FSO) system was presented, considering different atmospheric turbulence models and pointing errors. 

In light of the above, there is a research gap in the exact performance analysis of wireless systems requiring statistical characterization for the sum of multivariate Fox-H function. It should be noted that multi-parameter generalized fading models   contain multiple mathematical functions requiring more than single-variate Fox-H representation. Interestingly, the Rician fading model, a popular model for line-of-sight propagation in RIS-assisted systems, needs a bivariate Fox-H function for the statistics of the product of two-channel fading coefficients. Most of the existing literature avoids the intricacy of analysis by representing one of the functions (mainly the underlying Bessel functions) using a convergent infinite series. Further, the present study on the ergodic capacity employ approximations \cite{trigui2020_fox, du2020_thz, chapala2021THz, chapala2021unified,chapala2022reconfigurable},  even when the statistics of resultant signal-to-noise ratio (SNR) consist of the multivariate Fox-H obtained from the sum of single-variate Fox-H functions.

In this paper, we propose a novel approach to develop the distribution of the sum of independent and non-identically distributed random variables represented using multivariate Fox's H-function.  We also present a general framework for exact analysis of the  ergodic capacity when the multivariate Fox-H function characterizes SNR statistics.  We conduct an exact performance analysis, focusing on outage probability and ergodic capacity, with a specific application to RIS-assisted communication over Rician fading channels with phase errors. Through computer simulations, we demonstrate the superior accuracy of our approach compared to approximations and infinite-series expansions across a range of practical scenarios, providing a more accurate  performance evaluation.

\subsection{ Related Works}
 Initial research focused on applying the central limit theorem (CLT) to approximate the analysis of various fading models for RIS-based systems \cite{Kudathanthirige2020, Qin2020, Ferreira2020, Selimis2021, Salhab2021, Wang_2020_outage}. In \cite{Chandradeep2021}, the authors considered the probability density function (PDF) of double-Rician distributions using infinite series involving Bessel functions to analyze the ergodic capacity and average symbol-error rate (SER) performance of RIS-assisted systems over Rician fading for mmWave communication. 
 
  Recent studies have approximated the performance of RIS systems under different channel fading models in the presence of phase error \cite{Kudathanthirige2020, Qian_RIS_phase, LiDong_RIS_phase, wang_RIS_phase, Peng_RIS_phase, waqar2021performance, Tegos_RIS_phase,Bao2022,chapala2022reconfigurable}. For instance, in \cite{Kudathanthirige2020}, \cite{Qian_RIS_phase}, \cite{LiDong_RIS_phase}, and \cite{wang_RIS_phase}, the authors utilized the central limit theorem (CLT) to analyze the performance of RIS systems over Rayleigh fading channels with phase errors. In \cite{Badiu_RIS_phase}, an arbitrary fading model was approximated using the Nakagami-$m$ distribution to develop a performance analysis of RIS-assisted transmissions with phase error. The effect of diversity order on system performance was investigated in \cite{Peng_RIS_phase} under the constraint of finite quantization levels for the phase error. Additionally, performance bounds were derived in \cite{waqar2021performance} for RIS-assisted wireless systems with symmetric Rician fading, considering both fading channels and phase error at the RIS module. In \cite{Bao2022}, the distribution of the product of two Rician distributions with phase errors  due to the imperfect channel state information (CSI) was expressed in terms of infinite series containing a Meijer's G-function. 
  
  In our previous study \cite{chapala2022reconfigurable}, performance analysis of a RIS-assisted vehicular communication system considering generalized fading channels, random phase error, and combined signals from RIS reflections and direct transmissions was presented. The analysis incorporated generalized-$K$ shadowed fading for the direct link, asymmetrical channels with $\kappa$-$\mu$ distribution for the first link, double generalized Gamma (dGG) distribution for the second link, and a statistical random way point (RWP) model for the moving destination. To the authors' best knowledge, there is no exact analysis for the RIS-assisted wireless systems over  fading channels with/without phase errors when the underlying density and distribution function contains multiple mathematical functions.
\section{Statistical Results on  Multi-Variate Fox-H function}
 Let us consider the PDF $f_{X_{i}}(x)$ of a random variable $X_i$ in terms of $M$-variate ($M\geq 1$) Fox-H function as
\begin{flalign}\label{eq:multvar_pdf}
&f_{X_{i}}(x) = \psi_{i} x^{a_{i}-1} 	H_{p,q:p_{1},q_{1};\cdots;p_{M},q_{M}}^{0,n:m_{1},n_{1};\cdots;m_{M},n_{M}} \nonumber \\ &\bigg[\begin{array}{c} \{\zeta_{i,k}^{} x^{a_{i,k}}\}_{k=1}^{M} \end{array} \big\vert \begin{array}{c}\{(\alpha_{i,j}:\{A_{i,j}^{(k)}\}_{k=1}^{M})\}_{j=1}^{p}:\{U_{k}\}_{k=1}^{M}\\ \{(\beta_{i,j}:\{B_{i,j}^{(k)}\}_{k=1}^{M})\}_{j=1}^{q}:\{V_{k}\}_{k=1}^{M} \end{array}\bigg] 
\end{flalign}
where $U_{k}=\{(c_{i,j}^{(k)},C_{i,j}^{(k)})\}_{j=1}^{p_{k}}$ and $V_{k}=\{(d_{i,j}^{(k)},D_{i,j}^{(k)})\}_{j=1}^{q_{k}}$. In the following, we develop  statistical results on the sum of  random variable defined as $X_{ }=\sum_{i=1}^{N}X_{i}$:
\begin{my_theorem}\label{th:sum_mult_pdf_ris}
	The PDF and CDF of $X_{ }=\sum_{i=1}^{N}X_{i}$, where $X_i$ is distributed according to $M$-variate Fox-H function are given in terms of $MN$-variate Fox-H functions, as represented in  \eqref{eq:sum_mult_pdf} and \eqref{eq:sum_mult_cdf}, respectively.
	%	\small 
	\begin{figure*}
		\begin{eqnarray}\label{eq:sum_mult_pdf}
		&f_{X_{ }}(x) = \prod_{i=1}^{N} \psi_{i} x^{a_{i}-1} 	H_{N(p+1),Nq+1:p_{1},q_{1};\cdots;p_{M},q_{M};\cdots;p_{1},q_{1};\cdots;p_{M},q_{M}}^{0,N(n+1):m_{1},n_{1};\cdots;m_{M},n_{M};\cdots;m_{1},n_{1};\cdots;m_{M},n_{M}} \nonumber \\ &\bigg[\begin{array}{c} \{\{\zeta_{i,k}^{} x^{a_{i,k}}\}_{k=1}^{M}\}_{i=1}^{N} \end{array} \big\vert \begin{array}{c}V_{1}:\{\{\{(c_{i,j}^{(k)},C_{i,j}^{(k)})\}_{j=1}^{p_{k}}\}_{k=1}^{M}\}_{i=1}^{N}\\ V_{2},(1-\sum_{i=1}a_{i}:\{\{a_{i,k}\}_{k=1}^{M}\}_{i=1}^{N}):\{\{\{(d_{i,j}^{(k)},D_{i,j}^{(k)})\}_{j=1}^{q_{k}}\}_{k=1}^{M}\}_{i=1}^{N} \end{array}\bigg]  
		\end{eqnarray}
		\begin{eqnarray}\label{eq:sum_mult_cdf}
		&F_{X_{ }}(x) = \prod_{i=1}^{N} \psi_{i} x^{a_{i}} 	H_{N(p+1),Nq+1:p_{1},q_{1};\cdots;p_{M},q_{M};\cdots;p_{1},q_{1};\cdots;p_{M},q_{M}}^{0,N(n+1):m_{1},n_{1};\cdots;m_{M},n_{M};\cdots;m_{1},n_{1};\cdots;m_{M},n_{M}} \nonumber \\ &\bigg[\begin{array}{c} \{\{\zeta_{i,k}^{} x^{a_{i,k}}\}_{k=1}^{M}\}_{i=1}^{N} \end{array} \big\vert \begin{array}{c}V_{1}:\{\{\{(c_{i,j}^{(k)},C_{i,j}^{(k)})\}_{j=1}^{p_{k}}\}_{k=1}^{M}\}_{i=1}^{N}\\ V_{2},(-\sum_{i=1}a_{i}:\{\{a_{i,k}\}_{k=1}^{M}\}_{i=1}^{N}):\{\{\{(d_{i,j}^{(k)},D_{i,j}^{(k)})\}_{j=1}^{q_{k}}\}_{k=1}^{M}\}_{i=1}^{N} \end{array}\bigg]  
		\end{eqnarray}
		where $V_{1}=\{(1-a_{i}:\{a_{i,k}\}_{k=1}^{M},0,0,\cdots,0,0),\{(\alpha_{i,j}:\{A_{i,j}^{(k)}\}_{k=1}^{M},0,0,\cdots,0,0)\}_{j=1}^{p}\}_{i=1}^{N}$ and $V_{2}=\{\{(\beta_{i,j}:\{B_{i,j}^{(k)}\}_{k=1}^{M},0,0,\cdots,0,0)\}_{j=1}^{q}\}_{i=1}^{N}$.
		\hrule
	\end{figure*}
\end{my_theorem}
\begin{IEEEproof} 
	The proof is presented in Appendix A.
\end{IEEEproof}
The results in Theorem 1 can be used to derive exact statistical performance of wireless systems requiring sum of more than single-variate Fox-H function. 

Next, we provide a general framework to analyze the ergodic rate and average SNR performance of wireless systems when the statistics of the resultant SNR is depicted by the multivariate Fox-H function, as given in \eqref{eq:multvar_pdf}. This situation arises even when the statistics of the individual channel can be modeled by single-variate requiring multivariate Fox-H function representation for the sum. In general, denoting the random variable $X_i$, we  define an integral of the following form:

\begin{equation}\label{eq:gen_cap2}
I_g = \int_{0}^{\infty} g(x)  f_{X_	i} (x) dx 
\end{equation}
where $g(x)=\log_2 (1 + x)$ can be used for ergodic capacity analysis and $g(x)=x^{n}$ for $n^{th}$-moment of SNR, if the random variable $X_i$ denotes the SNR of the system. The existing literature \cite{trigui2020_fox, du2020_thz, chapala2021THz, chapala2021unified,chapala2022reconfigurable} use the final value theorem (FLT) to find the limiting value of the inner integral of the Mellin-Barnes integral representation to solve \eqref{eq:gen_cap2}.  It should be noted  that the form of  $g(x)=x^pe^{-qx}$ (where $p$ and $q$ are modulation constants) used for the average bit-error-rate (BER) analysis has been derived in previous literature.
 \begin{my_proposition}\label{th:gen_capacity2}
 	Assuming $X_i$ to be an $M$-variate Fox-H function, then $I_g $ in \eqref{eq:gen_cap2} can be represented in a closed-form using $M+1$-variate Fox-H function if a converging function $C(x)$ is introduced into $I_g$ resulting into an increase of $1$-variate due to $C(x)$.  
 	
 \end{my_proposition}
  \begin{proof}
  We can express \eqref{eq:gen_cap2} in the following manner:
 \begin{equation}\label{eq:gen_cap4}
 I_g = \int_{0}^{\infty} C(x)g(x) C^{-1}(x)f_{X_i}(x) dx
 \end{equation}
 Here, the function $C(x)$ has been deliberately selected to satisfy the condition $C(x) C^{-1}(x) = 1$, ensuring the convergence of the inner integral. Additionally, this representation is in terms of the Gamma function for the Fox-H representation. The presence of $C^{-1}(x)$  increments the variate from $M$ to $M+1$.
 \end{proof}
 
 \begin{my_proposition}\label{th:gen_capacity}
Assuming $X_i$ to be an $M$-variate Fox-H function, then $I_g $ in \eqref{eq:gen_cap2} can be represented in a closed-form using $M+2$-variate Fox-H function if $g(x)=\log_2(1+x)$ and $M+1$-variate Fox-H function if $g(x)=x^n$.

\end{my_proposition}
 \begin{proof}
 Using \eqref{eq:multvar_pdf} in \eqref{eq:gen_cap2} , and applying the   definition of multivariate Fox-H function, the inner integral becomes $I=\int_{0}^{\infty} (x)^{\sum_{k=1}^{M}a_{i,k}s_{i,k}+a_{i}-1} g(x) dx$ which does not converge. The previous studies \cite{trigui2020_fox, du2020_thz, chapala2021THz, chapala2021unified,chapala2022reconfigurable} used the FLT to find a limiting value of $I$. 
 	 	
 	 	We provide a simple yet novel approach to find an exact  expression of $I$. We introduce a converging factor $e^{-x}$ in $I$ with an  increase of a variate $M$ to $M+1$ in  Mellin-Barnes integral \eqref{eq:gen_cap_4} as $e^{x}=G_{0,1}^{1,0}\big[\begin{array}{c} -x\end{array} \big\vert \begin{array}{c}-\\0\end{array}\big]=\frac{1}{2\pi\J}\int_{\mathcal{L}} (-x)^{s_{M+2}} \Gamma(-s_{M+2}) ds_{M+2}$. 
   Thus, the inner integral $I=\int_{0}^{\infty} e^{-x} (x)^{\sum_{k=1}^{M}a_{i,k}s_{i,k}+a_{i}+s_{M+1}-1} g(x) dx $ can be easily computed using standard mathematical formulae considering different $g(x)$. If $g(x)=x^n$, the inner integral $I$ can be represented in terms of Gamma functions. However, if 
 $g(x)=\log_2(1+x)$, we need an increase in the variate by employing $\ln (1 + x) = \frac{1}{2\pi \J} \int_{\mathcal{L}} \frac{\Gamma(1-s_{M+2})\Gamma(s_{M+2})^{2}}{\Gamma(1+s_{M+2})} \big(x\big)^{s_{M+2}} \diff s_{M+2}$.
 	
\end{proof}
We can use the results of Theorem 1 and Theorem 2 to analyze the exact performance of wireless systems considering intricate fading models characterized by multiple mathematical functions. In what follows, we take an example of  RIS-assisted communication over Rician fading channels with phase errors to conduct an exact performance analysis.

\section{ Application to RIS-Assisted Wireless  System Over Rician Fading}
Consider a transmission model where a  source  communicates to a  destination  through an RIS with $N$ reflecting elements. Thus, the received signal at destination through RIS is given as
\begin{equation}\label{model_1}
	y = \sqrt[]{P_{t}} x \sum_{i=1}^{N} h_{l,i} \lvert h_{i}^{}\rvert  g_{l,i}^{}  \lvert g_{i}^{}\rvert e^{\J\phi_{i}}  + v
\end{equation}
Here, $P_{t}$ represents the transmission power, while $x$ stands for the unit power information-bearing signal. The channel gain from the source to the RIS is denoted as $h_{l,i}$, and the channel gain from the RIS to the destination is represented by $g_{l,i}$. Further, $\lvert h_{i}^{}\rvert$ and $\lvert g_{i}^{}\rvert$ correspond to the channel fading coefficients between the source and the $i$-th element of the RIS, and between the $i$-th RIS element and the destination, respectively. Furthermore, $\phi_{i}$ indicates the residual phase error for the $i$-th RIS element due to imperfect phase compensation, while $v$ accounts for the additive Gaussian noise with a mean of zero and variance $\sigma_v^2$.

We assume the channel coefficient $\lvert h_{i}^{}\rvert$  from the source to the RIS distributed as the Rician fading with shape parameter $K_{i,1}$ and scale parameter $\Omega_{i,1}$. Similarly, the channel coefficient $\lvert g_{i}^{}\rvert$ from the RIS to the destination  is Rician distributed with parameters $K_{i,2}$ and $\Omega_{i,2}$. The PDF of channel fading $\lvert h_{i}^{}\rvert$ is given as \cite{digital2005}
\begin{eqnarray}\label{eq:hi_pdf}
	&f_{\lvert h_{i}^{}\rvert}(x) = \frac{2 (1+K_{i,1})^{} x^{}}{\Omega_{i,1} {\rm exp}( K_{i,1})} {\rm exp}\Big[- \frac{(1+K_{i,1})}{\Omega_{i,1}}x^2\Big] \nonumber\\&I_{0}\Big[2 \sqrt{\frac{K_{i,1}(1+K_{i,1})}{\Omega_{i,1}}} x\Big]
\end{eqnarray} 
where $K_{i,1}$ is the ratio of power of the dominant components and the total power of scattered waves and $I_{v}(x)$ is the modified Bessel function of the first kind of $v$-order. The shape parameter and scale parameter can be, respectively written as $K_{i} = \frac{S_{i}}{2\sigma_{i}^{2}}$ and $\Omega_{i} = S_{i}^{2}+2\sigma_{i}^{2}$, where $S_{i}^{2}$ and $2\sigma_{i}^{2}$  represent the power of LOS and non-LOS components. Similar to the PDF of $|h_{i}|$ in \eqref{eq:hi_pdf}, we can express the PDF of the channel fading $\lvert g_{i}^{}\rvert$ with parameters $K_{i,2}$ and $\Omega_{i,2}$.

The origin of phase errors can arise from two sources: quantized phase compensation or errors in channel estimation. Within the literature, three models are presented to describe phase errors: Gaussian, generalized uniform, and uniform random variables (RVs) as outlined in \cite{Trigui_phasenoise}. We opt to utilize a generalized uniform distribution, denoted as $\phi_{i}\sim\mathcal{U}(-q\pi,q\pi)$, to characterize the effects of quantization noise. Here, $q=2^{-L}$ with $L\geq1$ representing the number of quantization bits.

To analyze the RIS performance in \eqref{model_1}, we require the sum of the product of Rician fading coefficients. We employ the approach as presented in \cite{trigui2020_fox, chapala2022reconfigurable} to get the PDF of $Z_{i}=Z_{hg,i}e^{\J\phi_{i}}=\lvert h_{i}^{}\rvert \lvert g_{i}^{}\rvert e^{\J\phi_{i}}$ as
\begin{flalign}\label{eq:pdf_z_i}
&f_{Z_{i}^{}}(x) = \frac{x^{-1}}{4{\rm exp}( K_{i,1}+K_{i,2})}H_{2,0:1,1;0,2;0,2}^{0,2:0,0;1,0;1,0}\nonumber\\&\hspace{-4mm}\bigg[\begin{array}{c}x^{-1} \sqrt{\zeta_{i}} \\ \sqrt{-K_{i,1}} \\ \sqrt{-K_{i,2}}\end{array}  \big\vert \begin{array}{c} (0:\frac{1}{2},\frac{1}{2},0),(0:\frac{1}{2},0,\frac{1}{2}):(1,q);-;- \\ -:(0,q);(0,\frac{1}{2}),(0,\frac{1}{2});(0,\frac{1}{2}),(0,\frac{1}{2})\end{array}\bigg]   
\end{flalign}

The PDF of $Z_{i}$ in \eqref{eq:pdf_z_i} is a bivariate Fox's H-function and there is no results on the PDF of its sum in the literature.
Next, we use the results of Theorem 1 to develop statistical analysis for $Z_{\rm RIS}=\sum_{i=1}^{N}Z_{i}$ to analyze the performance of  RIS-assisted communication over Rician fading channels:
\begin{my_lemma}\label{lemma:sum_mult_pdf_ris}
	The PDF and CDF of the effective channel through RIS, $Z_{\rm RIS}=\sum_{i=1}^{N}Z_{i}$ are given by \eqref{eq:ris_pdf} and \eqref{eq:ris_cdf} respectively.
	%	\small 
\begin{figure*}
	\begin{eqnarray}\label{eq:ris_pdf}
		&f_{Z_{\rm RIS}}(x) = \prod_{i=1}^{N} \frac{x^{-1}}{4{\rm exp}( K_{i,1}+K_{i,2})}  	H_{3N,1:1,1;0,2;0,2;\cdots;1,1;0,2;0,2}^{0,3N:0,0;1,0;1,0;\cdots;0,0;1,0;1,0} \nonumber \\ &\bigg[\begin{array}{c} \{x^{-1} \sqrt{\zeta_{i}}_{},\sqrt{-K_{i,1}},\sqrt{-K_{i,2}}\}_{i=1}^{N} \end{array} \big\vert \begin{array}{c}V_{1}:(1,q);-;-;\cdots;(1,q);-;-\\ V_{2},(1:\{-1,0,0\}_{i=1}^{N}):\{(0,q);(0,\frac{1}{2}),(0,\frac{1}{2});(0,\frac{1}{2}),(0,\frac{1}{2})\}_{i=1}^{N} \end{array}\bigg]  
	\end{eqnarray}
	\begin{eqnarray}\label{eq:ris_cdf}
		&F_{Z_{\rm RIS}}(x) = \prod_{i=1}^{N} \frac{1}{4{\rm exp}( K_{i,1}+K_{i,2})}  	H_{3N,1:1,1;0,2;0,2;\cdots;1,1;0,2;0,2}^{0,3N:0,0;1,0;1,0;\cdots;0,0;1,0;1,0} \nonumber \\ &\bigg[\begin{array}{c} \{x^{-1} \sqrt{\zeta_{i}}_{},\sqrt{-K_{i,1}},\sqrt{-K_{i,2}}\}_{i=1}^{N} \end{array} \big\vert \begin{array}{c}V_{1}:(1,q);-;-;\cdots;(1,q);-;-\\ (0:\{-1,0,0\}_{i=1}^{N}):\{(0,q);(0,\frac{1}{2}),(0,\frac{1}{2});(0,\frac{1}{2}),(0,\frac{1}{2})\}_{i=1}^{N} \end{array}\bigg]  
	\end{eqnarray}
	where $V_{1}=\{(1:\{-1,0,0\}_{i}^{},0,0,\cdots,0,0),\{(0:\{\frac{1}{2},\frac{1}{2},0\}_{i}^{},0,0,\cdots,0,0),(0:\{\frac{1}{2},0,\frac{1}{2}\}_{i}^{},0,0,\cdots,0,0)\}_{i=1}^{N}$.
	\hrule
\end{figure*}
\end{my_lemma}
\begin{IEEEproof} 
	A straight forward application of Theorem 1 completes the proof.
\end{IEEEproof}

Next, we use the statistical results of Lemma 1 to develop outage probability of the considered system.
\subsection{Outage Probability}
Outage probability is a performance metric to characterize the impact of fading in a communication system.  For the received signal through RIS, an expression for the  SNR is given as $\gamma_{RIS} = \gamma_{0, RIS} Z_{\rm RIS}^{2}$,	where $\gamma_{0,RIS}=\frac{h_{l_{}}^{2} g_{l_{}}^{2} P_{t}}{\sigma_{v}^{2}}$  represents average SNR for RIS transmission.

	\begin{figure*}
	\begin{eqnarray}\label{eq:rician_cap}
	&\bar{\eta}_{\rm RIS} =    \prod_{i=1}^{N} \frac{\log_2(e)}{8{\rm exp}( K_{i,1}+K_{i,2})}  	H_{3N+1,1:1,1;0,2;0,2;\cdots;1,1;0,2;0,2;2,2;0,1}^{0,3N+1:0,0;1,0;1,0;\cdots;0,0;1,0;1,0;1,2;1,0} \nonumber \\ &\bigg[\begin{array}{c} \{\sqrt{\gamma_{0,RIS}} \sqrt{\zeta_{i}}_{},\sqrt{-K_{i,1}},\sqrt{-K_{i,2}}\}_{i=1}^{N},1,-1 \end{array} \big\vert \begin{array}{c}V_{1}:(1,q);-;-;\cdots;(1,q);-;-;(1,1),(1,1);-\\ (1:\{-1,0,0\}_{i=1}^{N},0,0):V_{2};(1,1),(0,1);(0,1) \end{array}\bigg]
	\end{eqnarray}
	where $V_{1}=\{(1:\{-1,0,0\}_{i}^{},0,0,\cdots,0,0,0,0),(0:\{\frac{1}{2},\frac{1}{2},0\}_{i}^{},0,0,\cdots,0,0,0,0),(0:\{\frac{1}{2},0,\frac{1}{2}\}_{i}^{},0,0,\cdots,0,0,0,0)\}_{i=1}^{N},(1:\{-1/2,0,0\}_{1}^{N},1,1)$ and $V_{2}=\{(0,q);(0,\frac{1}{2}),(0,\frac{1}{2});(0,\frac{1}{2}),(0,\frac{1}{2})\}_{i=1}^{N}$. 
	\hrule
\end{figure*}

Using the definition of the outage probability  $ P_{\rm out}=Pr(\gamma \le \gamma_{\rm th})$, where $\gamma_{\rm th}$  is the threshold SNR, an exact expression for the outage probability is given by substituting $x$ with $\gamma_{\rm th}$ in  \eqref{eq:ris_cdf} with a random-variable transformation i.e,  $F_{\gamma_{RIS}}(\gamma_{\rm th}) = F_{Z_{\rm RIS}}(\sqrt{\frac{\gamma_{\rm th}}{\gamma_{0, RIS}}})$.

To derive the diversity order of the considered system, we first compute the dominant pole as $p_{i,k}=\min{\{\{\frac{D_{i,j}^{(k)}}{d_{i,j}^{(k)}}\}_{j=1}^{m_{k}}\}_{k=1}^{M}}$ for the multivariate CDF expression of $F_{\gamma_{RIS}}(\gamma_{\rm th})$. We find the residue at this dominant pole to compute the outage probability asymptotically at high SNR as \eqref{eq:out_asymp}. We express the outage probability as $P_{\rm out}^{\infty} \propto \gamma_{0, RIS}^{-G_{\rm out}}$ which gives the outage diversity order as $G_{\rm out}= \sum_{i=1}^{N}(\frac{a_{i}+\sum_{k=1}^{M}a_{i,k}p_{i,k}}{2})$. For the considered Rician fading model the outage diversity order becomes $G_{\rm out}=N$, which is consistent of previous results \cite{trigui2020_fox}.

\begin{eqnarray}\label{eq:out_asymp}
	&P_{\rm out}^{\infty} \approx \prod_{i=1}^{N} \psi_{i} (\frac{1}{2\pi \J})^{M} \prod_{k=1}^{M} (\zeta_{i,k}^{})^{p_{i,k}}  \nonumber \\&\hspace{-4mm}\frac{\prod_{j=1}^{n}\Gamma(1-\alpha_{i,j}+\sum_{k=1}^{M}A_{i,j}^{(k)}p_{i,k})\Gamma(\sum_{k=1}^{M}a_{i,k}p_{i,k}+a_{i})}{\prod_{j=n+1}^{p}\Gamma(\alpha_{i,j}-\sum_{k=1}^{M}A_{i,j}^{(k)}p_{i,k})\prod_{j=1}^{q}\Gamma(1-\beta_{i,j}+\sum_{k=1}^{M}B_{i,j}^{(k)}p_{i,k})}\nonumber \\ &\hspace{-4mm} \prod_{k=1}^{M}\frac{\prod_{j=1}^{m_{k}}\Gamma(d_{i,j}^{(k)}-D_{i,j}^{(k)}p_{i,k})\prod_{j=1}^{n_{k}}\Gamma(1-c_{i,j}^{(k)}+C_{i,j}^{(k)}p_{i,k})}{\prod_{j=n_{k}+1}^{p_{k}}\Gamma(c_{i,j}^{(k)}-C_{i,j}^{(k)}p_{i,k})\prod_{j=m_{k}+1}^{q_{k}}\Gamma(1-d_{i,j}^{(k)}+D_{i,j}^{(k)}p_{i,k})} \nonumber \\& \hspace{-4mm}\frac{1}{\Gamma(\sum_{i=1}^{N}(\sum_{k=1}^{M}a_{i,k}p_{i,k})+a_{i}+1)} (\sqrt{\frac{\gamma}{\gamma_{0, RIS}}})^{a_{i}+\sum_{k=1}^{M}a_{i,k}p_{i,k}}
\end{eqnarray}

\subsection{Ergodic Capacity}\label{sec:capacity}
Ergodic capacity for an RIS-assisted system is an important metric to an estimate of the average throughput, as defined in the following
\begin{equation}\label{eq:gen_cap}
	\bar{\eta} 
	= \int_{0}^{\infty} \log_2 (1 + \gamma)  f_{\gamma_{RIS}} (\gamma) d\gamma 
\end{equation}
Applying a transformation on \eqref{eq:ris_pdf} to get $f_{\gamma_{RIS}}(\gamma) = \frac{1}{2\sqrt{\gamma_{0, RIS} \gamma}}f_{Z_{\rm RIS}}(\sqrt{\frac{\gamma}{\gamma_{0, RIS}}})$, and with $\ln (1 + \gamma) = \frac{1}{2\pi \J} \int_{\mathcal{L}} \frac{\Gamma(1-s_{3N+1})\Gamma(s_{3N+1})^{2}}{\Gamma(1+s_{3N+1})} \big(\gamma\big)^{s_{3N+1}} \diff s_{3N+1}$, and applying 
the   definition of multivariate Fox-H function, 
	\begin{eqnarray}\label{eq:gen_cap_4}
	&\bar{\eta} =     \frac{\psi_{i}}{2} (\sqrt{\frac{1}{\gamma_{0}}})^{a_{i}} (\frac{1}{2\pi \J})^{M} (\frac{1}{2\pi \J}) \nonumber \\&\int_{\mathcal{L}_{i,1}} \cdots \int_{\mathcal{L}_{i,M}} \int_{\mathcal{L}} \prod_{k=1}^{M} (\zeta_{i,k}^{})^{s_{i,k}} (\sqrt{\frac{1}{\gamma_{0}}})^{a_{i,k}s_{i,k}} (-1)^{s_{M+2}} \nonumber \\ &\frac{\prod_{j=1}^{n}\Gamma(1-\alpha_{i,j}+\sum_{k=1}^{M}A_{i,j}^{(k)}s_{i,k})}{\prod_{j=n+1}^{p}\Gamma(\alpha_{i,j}-\sum_{k=1}^{M}A_{i,j}^{(k)}s_{i,k})\prod_{j=1}^{q}\Gamma(1-\beta_{i,j}+\sum_{k=1}^{M}B_{i,j}^{(k)}s_{i,k})}\nonumber \\ & \prod_{k=1}^{M}\frac{\prod_{j=1}^{m_{k}}\Gamma(d_{i,j}^{(k)}-D_{i,j}^{(k)}s_{i,k})\prod_{j=1}^{n_{k}}\Gamma(1-c_{i,j}^{(k)}+C_{i,j}^{(k)}s_{i,k})}{\prod_{j=n_{k}+1}^{p_{k}}\Gamma(c_{i,j}^{(k)}-C_{i,j}^{(k)}s_{i,k})\prod_{j=m_{k}+1}^{q_{k}}\Gamma(1-d_{i,j}^{(k)}+D_{i,j}^{(k)}s_{i,k})} \nonumber \\ & \frac{\Gamma(1-s_{M+1})\Gamma(s_{M+1})^{2}}{\Gamma(1+s_{M+1})} \int_{0}^{\infty} (\gamma)^{\frac{1}{2}(-\sum_{k=1}^{N}s_{k})+s_{3N+1}-1} d\gamma \nonumber\\&ds_{i,1} \cdots ds_{i,M}  \diff s_{M+1} ds_{M+2}
\end{eqnarray}
The inner integral $I=\int_{0}^{\infty} (\gamma)^{\frac{1}{2}(-\sum_{k=1}^{N}s_{k})+s_{3N+1}-1} d\gamma$ does not converge. We introduce a converging factor $C(\gamma)=e^{-\gamma}$ in $I$ with an  increase of a variate $3N+1$ to $3N+2$ in  Mellin-Barnes integral as $e^{\gamma}=G_{0,1}^{1,0}\bigg[\begin{array}{c} -\gamma\end{array} \big\vert \begin{array}{c}-\\0\end{array}\bigg]=\frac{1}{2\pi\J}\int_{\mathcal{L}} (-\gamma)^{s_{3N+2}} \Gamma(-s_{3N+2}) ds_{3N+2}$. 
Thus, using the inner integral $I=\int_{0}^{\infty} e^{-\gamma} (\gamma)^{\frac{1}{2}(-\sum_{k=1}^{N}s_{i,k})+s_{3N+1}+s_{3N+2}-1} d\gamma = \Gamma(\frac{1}{2}(-\sum_{k=1}^{N}s_{i,k})+s_{3N+1}+s_{3N+2})$ in \eqref{eq:gen_cap_4} with standard mathematical procedure of dealing multivariate Fox-H representation, we get \eqref{eq:rician_cap}.

\section{Simulation and Numerical Results}\label{sec:sim_num_res}
In this section, we validate the derived analytical expressions through numerical analysis and Monte Carlo simulations (averaged over $10^{8}$ channel realizations). We consider carrier frequency $f=6$\mbox{GHz}, transmit antenna gain $G_{T}=10$\mbox{dBi}, receive antenna gain $G_{R}=10$\mbox{dBi}, the distance from source to RIS, $d_{1}=20$\mbox{m} and from RIS to destination, $d_{2}=100$\mbox{m}. We vary Rician fading shape and scale parameters and we assume i.i.d channel from source to RIS and RIS to destination i.e., $K_{i,1}=K_{i,2}=K$, $\Omega_{i,1}=\Omega_{i,2}=\Omega{i}$ $\forall i$ to analyze the performance of the considered RIS system. A noise floor of $-74$\mbox{dBm} is considered over a $20$\mbox{MHz} channel bandwidth.

\begin{figure*}
	\centering
	\subfigure[PDF ]{\includegraphics[scale=0.35]{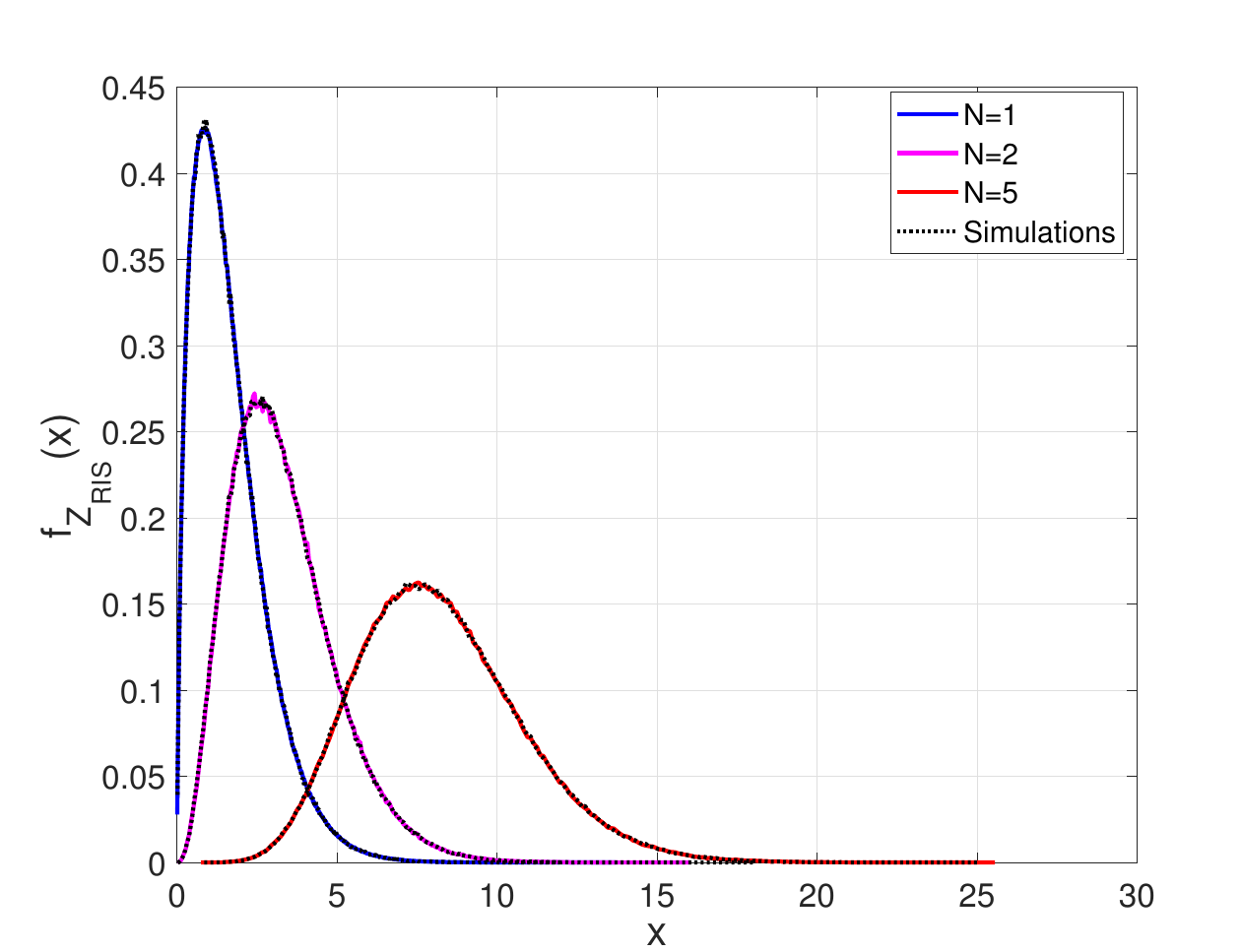 }} 
	\subfigure[CDF]{\includegraphics[scale=0.35]{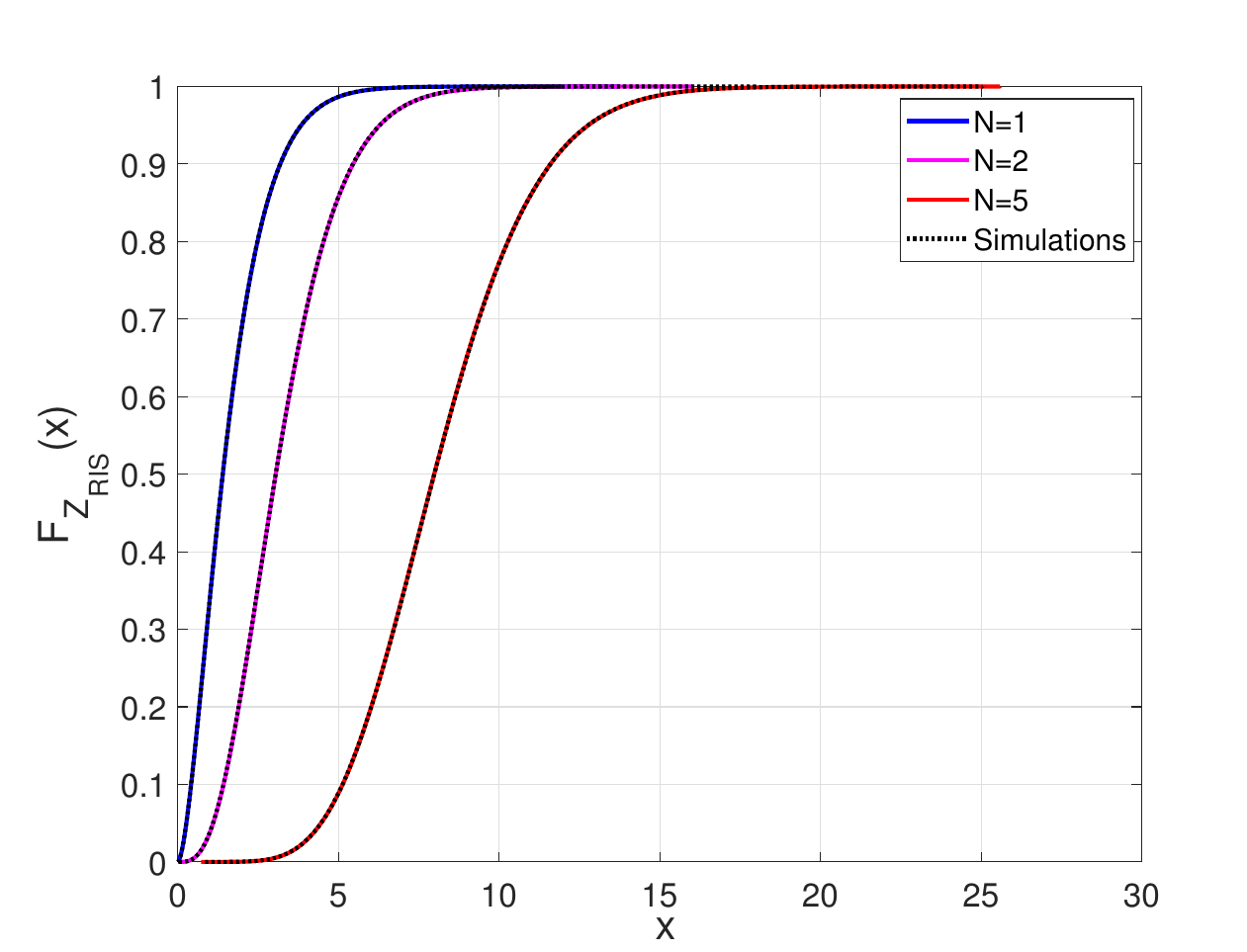 }}
		\caption{PDF and CDF of RIS-assisted channel over Rician fading with $K=1$ and $\sigma^2=0.5$.}
	\label{fig:pdf}
\end{figure*}

 In Fig.~\ref{fig:pdf}, we validate the PDF and CDF derived from Theorem 1, which pertain to the summation of the product of channel fading coefficients following a Rician distribution. The illustration demonstrates a close alignment between the numerical outcomes and simulation results. Additionally, it is observed that as the number of RIS elements increases, the channel's suitability for transmission becomes more favorable.

\begin{figure*}
	\subfigure[Outage probability for $N=5$.]{\includegraphics[scale=0.28]{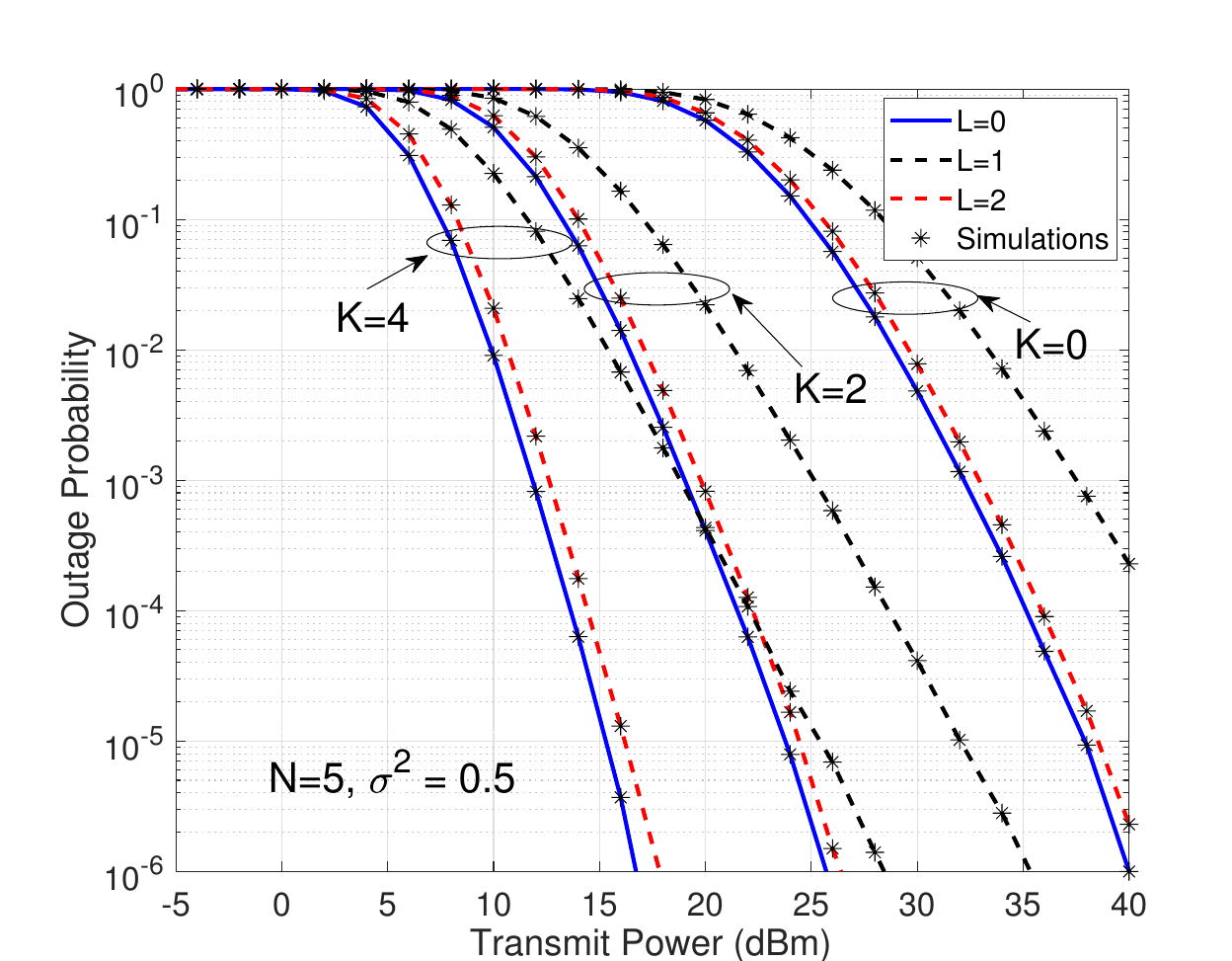}} 
	\subfigure[Outage probability for  $K=2$. ]{\includegraphics[scale=0.28]{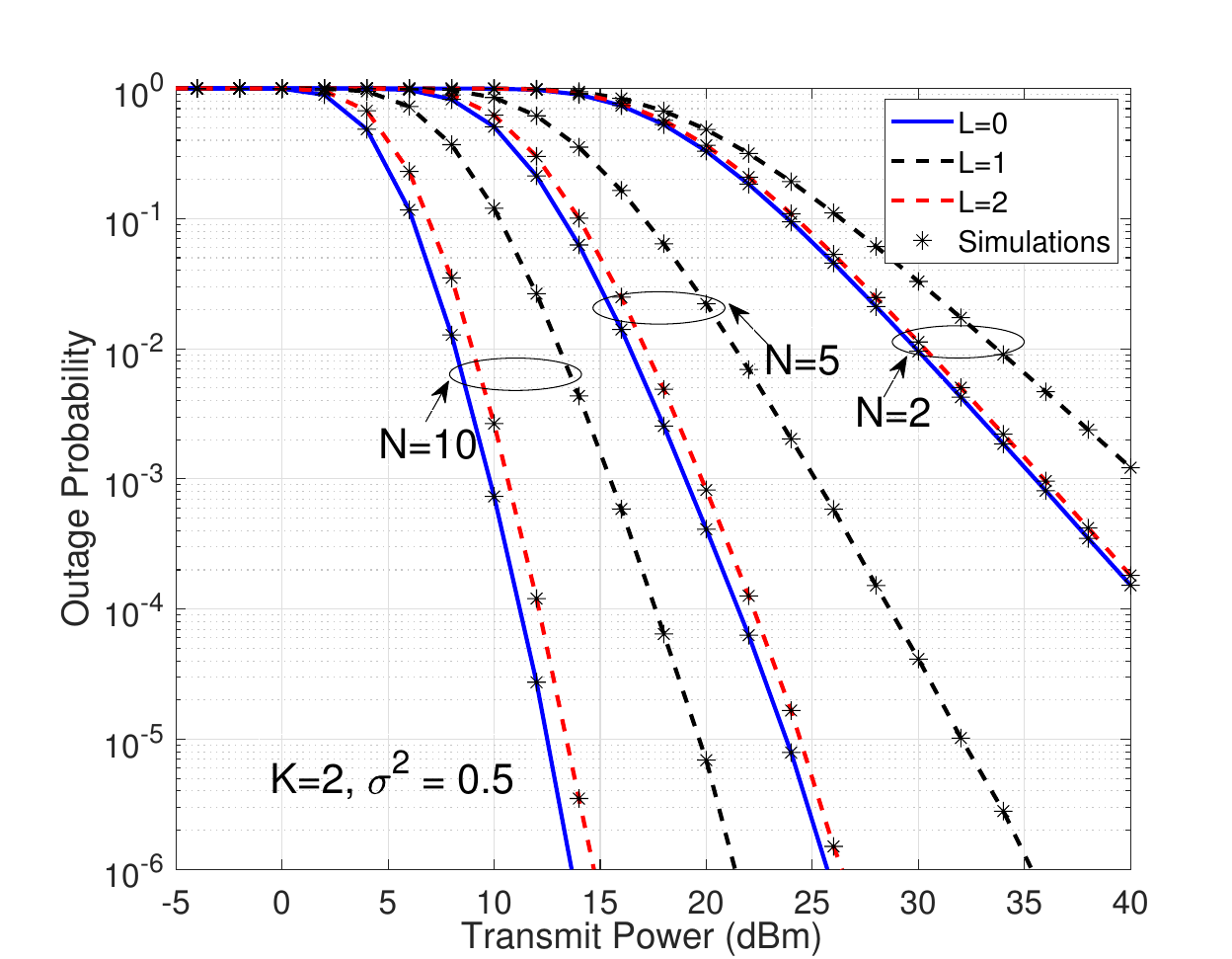}}
	\subfigure[Ergodic capacity for $K=2$. ]{\includegraphics[scale=0.28]{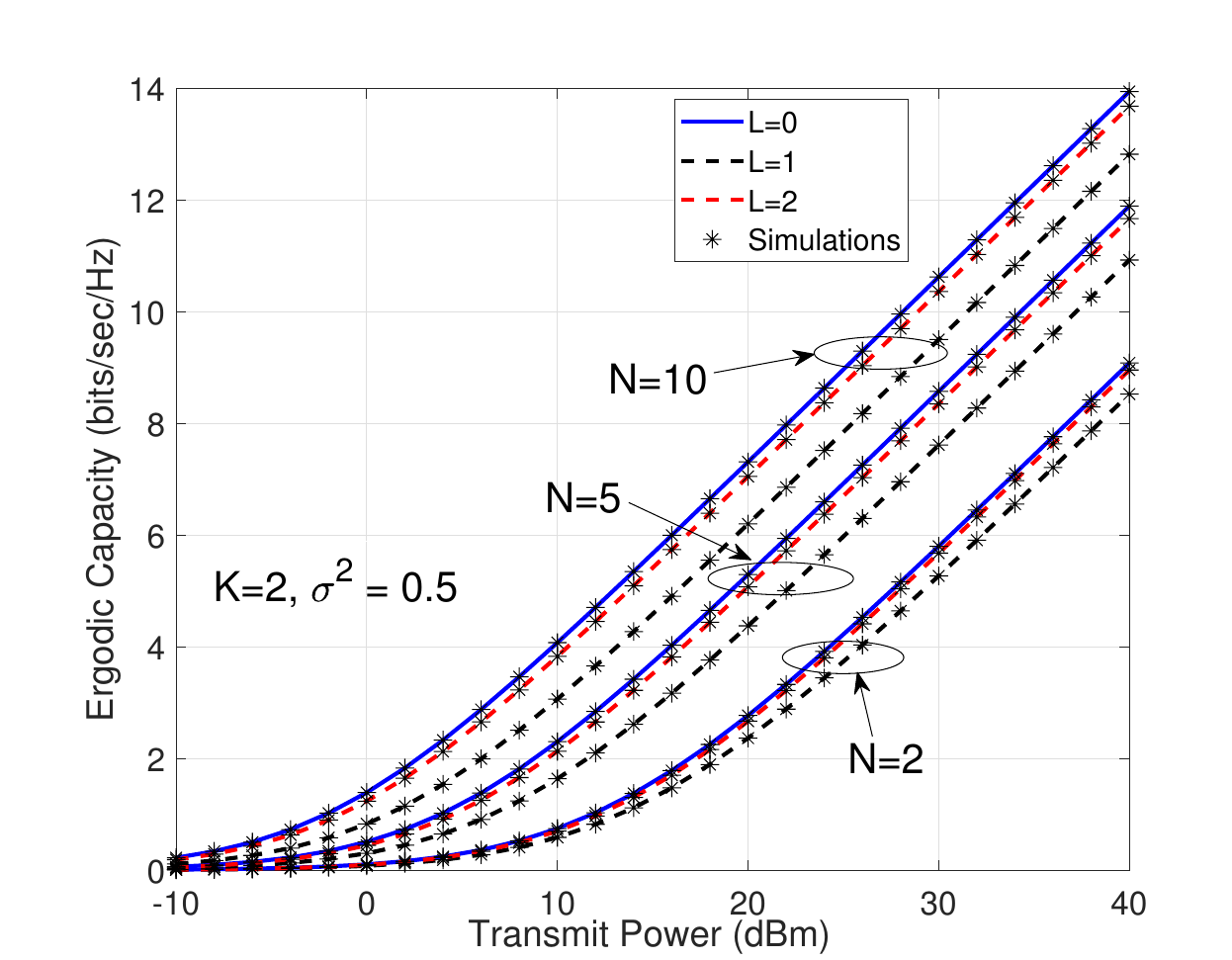}}
	\caption{Performance of  the RIS system Over Rician Fading.}
	\label{fig:outage_ris}
	%	\label{fig:ber_mob}
\end{figure*}

Fig.~\ref{fig:outage_ris}(a) shows the outage performance of RIS system over Rician fading for different values of shape parameter $K$ and for a given $N=5$, $\sigma_{i,1}^{2}=\sigma_{i,2}^{2}=1/2$ $\forall i$ and for two different quantization levels of phase error. We have used $\gamma_{\rm th}=6$\mbox{dB} to compute the outage probability. Here, the impact of Rician factor $K$ on system performance is depicted. It can be seen that the performance improves with an increase in $K$ due to increase in the ratio of power of LOS  to non-LOS components. We have also plotted RIS performance with $1$-bit ($L=1$) and $2$-bit ($L=2$) quantization to represent phase error at RIS and compared it without phase error ($L=0$). With $L=1$, there is a $5\mbox{dBm}$ loss in transmit power to achieve a desired outage performance of $10^{-3}$. However, with an increase in quantization bits $L=2$, the performance of RIS system improves and is close to the case of no phase error.   In Fig.~\ref{fig:outage_ris}(b), we plot the outage probability for different RIS elements $N$ and Rician factor $K=2$. Here, the outage performance improves with larger RIS size $N$. Further for a given $K$, increasing the RIS size gives the better outage performance. It means that the randomness introduced in the transmission channel  through the Rician factor $K$ can be compensated with appropriate selection of number of reflecting elements.

Finally, in Fig.~\ref{fig:outage_ris}(c), we plot the ergodic capacity for different RIS elements $N$ and Rician factor $K=2$. Here, the ergodic capacity improves with larger RIS size $N$. Further for a given $K$, increasing the RIS size gives the better capacity performance. The ergodic capacity  improves with increase in RIS size $N$ and for higher quantization level $L=2$. 

\section{Conclusions}
In this paper, we presented exact  closed-form expressions on the sum Multivariate Fox's H-functions and used the derived results to compute the exact density, distribution functions, and  outage performance of RIS assisted wireless communication system over Rician fading model.
The analysis that was proposed involved deriving the PDF and CDF of the resulting channel between a source and destination, with the RIS module incorporating multiple reflective elements. The performance of the system under consideration was showcased by developing exact and asymptotic expressions for the outage probability and ergodic capacity. Computer simulations were carried out to showcase the precision of our analysis in enhancing the evaluation of performance across a range of practically significant scenarios.
%\section*{Appendix A: PDF and CDF of Direct Link $h_i$}
%
%  \section*{Appendix B: PDF, CDF, and MGF of $Z_i$}
%
%
%\section*{Appendix C: PDF and CDF of $Z$}

\section*{Appendix A}
The PDF of the sum, $X=\sum_{i=1}^{N}X_{i}$ can be computed from its MGF as $f_{X_{ }}(z) = \mathcal{L}^{-1} \prod_{i=1}^{N} M_{X_i}(s)	$. We substitute the PDF of  $X_{i}$ given in \eqref{eq:multvar_pdf}, expand the M-multivariate Fox's H-function definition and interchange the order of integration to get
\begin{eqnarray}\label{eq:mgf_mult_zi_1}
&\hspace{-4mm}M_{X_i}(s) = \psi_{i} (\frac{1}{2\pi \J})^{M} \int_{\mathcal{L}_{i,1}} \cdots \int_{\mathcal{L}_{i,M}} \prod_{k=1}^{M} (\zeta_{i,k}^{})^{s_{i,k}}   \nonumber \\ &\frac{\prod_{j=1}^{n}\Gamma(1-\alpha_{i,j}+\sum_{k=1}^{M}A_{i,j}^{(k)}s_{i,k})}{\prod_{j=n+1}^{p}\Gamma(\alpha_{i,j}-\sum_{k=1}^{M}A_{i,j}^{(k)}s_{i,k})\prod_{j=1}^{q}\Gamma(1-\beta_{i,j}+\sum_{k=1}^{M}B_{i,j}^{(k)}s_{i,k})}\nonumber \\ & \prod_{k=1}^{M}\frac{\prod_{j=1}^{m_{k}}\Gamma(d_{i,j}^{(k)}-D_{i,j}^{(k)}s_{i,k})\prod_{j=1}^{n_{k}}\Gamma(1-c_{i,j}^{(k)}+C_{i,j}^{(k)}s_{i,k})}{\prod_{j=n_{k}+1}^{p_{k}}\Gamma(c_{i,j}^{(k)}-C_{i,j}^{(k)}s_{i,k})\prod_{j=m_{k}+1}^{q_{k}}\Gamma(1-d_{i,j}^{(k)}+D_{i,j}^{(k)}s_{i,k})} \nonumber \\&\hspace{-4mm}\int_{0}^{\infty} e^{-sx} x^{\sum_{k=1}^{M}a_{i,k}s_{i,k}+a_{i}-1} dx ds_{i,1} \cdots ds_{i,M}
\end{eqnarray}

Now, the inner integral in \eqref{eq:mgf_mult_zi_1} can be solved using the identity \cite[3.381.4]{integrals} as $\int_{0}^{\infty} e^{-sx} x^{\sum_{k=1}^{M}a_{i,k}s_{i,k}+a_{i}-1} dx = s^{-\sum_{k=1}^{M}a_{i,k}s_{i,k}-a_{i}} \Gamma(\sum_{k=1}^{M}a_{i,k}s_{i,k}+a_{i})$. %We substitute it back to get
%\begin{eqnarray}\label{eq:mgf_mult_zi_2}
%	&M_{X_i}(s) = \psi_{i} (\frac{1}{2\pi \J})^{M} \int_{\mathcal{L}_{i,1}} \cdots \int_{\mathcal{L}_{i,M}} \prod_{k=1}^{M} (\zeta_{i,k}^{}s^{-a_{i,k}})^{s_{i,k}}  \nonumber \\& \frac{\prod_{j=1}^{n}\Gamma(1-\alpha_{i,j}+\sum_{k=1}^{M}A_{i,j}^{(k)}s_{i,k})\Gamma(\sum_{k=1}^{M}a_{i,k}s_{i,k}+a_{i})}{\prod_{j=n+1}^{p}\Gamma(\alpha_{i,j}-\sum_{k=1}^{M}A_{i,j}^{(k)}s_{i,k})\prod_{j=1}^{q}\Gamma(1-\beta_{i,j}+\sum_{k=1}^{M}B_{i,j}^{(k)}s_{i,k})}\nonumber \\ & \prod_{k=1}^{M}\frac{\prod_{j=1}^{m_{k}}\Gamma(d_{i,j}^{(k)}-D_{i,j}^{(k)}s_{i,k})\prod_{j=1}^{n_{k}}\Gamma(1-c_{i,j}^{(k)}+C_{i,j}^{(k)}s_{i,k})}{\prod_{j=n_{k}+1}^{p_{k}}\Gamma(c_{i,j}^{(k)}-C_{i,j}^{(k)}s_{i,k})\prod_{j=m_{k}+1}^{q_{k}}\Gamma(1-d_{i,j}^{(k)}+D_{i,j}^{(k)}s_{i,k})} \nonumber\\&s^{-a_{i}}ds_{i,1} \cdots ds_{i,M}
%\end{eqnarray}

%We apply the definition of $M$-multivariate Fox's H-function to get
%\begin{eqnarray}\label{eq:mgf_mult_zi_final}
%	&M_{Z_i}(s) = \psi_{i} s^{-a_{i}} 	H_{p+1,q:p_{1},q_{1};\cdots;p_{M},q_{M}}^{0,n+1:m_{1},n_{1};\cdots;m_{M},n_{M}} \nonumber \\ &\bigg[\begin{array}{c} \{\zeta_{i,k}^{} s^{-a_{i,k}}\}_{k=1}^{M} \\ \zeta_{i,2}^{} s^{-a_{i,2}} \end{array} \big\vert \begin{array}{c}(1-a_{i}:\{a_{i,k}\}_{k=1}^{M}),\{(\alpha_{i,j}:\{A_{i,j}^{(k)}\}_{k=1}^{M})\}_{j=1}^{p}:\{\{(c_{i,j}^{(k)},C_{i,j}^{(k)})\}_{j=1}^{p_{k}}\}_{k=1}^{M}\\ \{(\beta_{i,j}:\{B_{i,j}^{(k)}\}_{k=1}^{M})\}_{j=1}^{q}:\{\{(d_{i,j}^{(k)},D_{i,j}^{(k)})\}_{j=1}^{q_{k}}\}_{k=1}^{M} \end{array}\bigg]  
%\end{eqnarray}

Then, the MGF of the sum is given as $M_{X}(s)=\prod_{i=1}^{N} M_{X_i}(s)$ and thus its PDF can be computed using
\begin{eqnarray}\label{eq:sum_mult_pdf_1}
&f_{X_{ }}(x) = \mathcal{L}^{-1} \prod_{i=1}^{N} M_{Z_i}(s) \nonumber \\ &= \prod_{i=1}^{N} \psi_{i} (\frac{1}{2\pi \J})^{M} \int_{\mathcal{L}_{i,1}} \cdots \int_{\mathcal{L}_{i,M}} \prod_{k=1}^{M} (\zeta_{i,k}^{})^{s_{i,k}} s^{-a_{i}} \nonumber \\& \frac{\prod_{j=1}^{n}\Gamma(1-\alpha_{i,j}+\sum_{k=1}^{M}A_{i,j}^{(k)}s_{i,k})\Gamma(\sum_{k=1}^{M}a_{i,k}s_{i,k}+a_{i})}{\prod_{j=n+1}^{p}\Gamma(\alpha_{i,j}-\sum_{k=1}^{M}A_{i,j}^{(k)}s_{i,k})\prod_{j=1}^{q}\Gamma(1-\beta_{i,j}+\sum_{k=1}^{M}B_{i,j}^{(k)}s_{i,k})}\nonumber \\ & \prod_{k=1}^{M}\frac{\prod_{j=1}^{m_{k}}\Gamma(d_{i,j}^{(k)}-D_{i,j}^{(k)}s_{i,k})\prod_{j=1}^{n_{k}}\Gamma(1-c_{i,j}^{(k)}+C_{i,j}^{(k)}s_{i,k})}{\prod_{j=n_{k}+1}^{p_{k}}\Gamma(c_{i,j}^{(k)}-C_{i,j}^{(k)}s_{i,k})\prod_{j=m_{k}+1}^{q_{k}}\Gamma(1-d_{i,j}^{(k)}+D_{i,j}^{(k)}s_{i,k})} \nonumber \\& (\frac{1}{2\pi \J} \int_{\mathcal{L}} s^{\sum_{i=1}^{N}-(\sum_{k=1}^{M}a_{i,k}s_{i,k})-a_{i}} e^{sx} ds) ds_{i,1} \cdots ds_{i,M}
\end{eqnarray}

We apply the identity \cite[8.315.1]{integrals} to solve the inner integral as
\begin{eqnarray}\label{eq:sum_mult_pdf_inner_int}
&\frac{1}{2\pi \J} \int_{\mathcal{L}} s^{\sum_{i=1}^{N}-(\sum_{k=1}^{M}a_{i,k}s_{i,k})-a_{i}} e^{sx} ds \nonumber\\&= \frac{x^{\sum_{i=1}^{N}(\sum_{k=1}^{M}a_{i,k}s_{i,k})+a_{i}-1}}{\Gamma(\sum_{i=1}^{N}(\sum_{k=1}^{M}a_{i,k}s_{i,k})+a_{i})} 
\end{eqnarray}

We use \eqref{eq:sum_mult_pdf_inner_int} in \eqref{eq:sum_mult_pdf_1} and apply the definition of N-multivariate Fox's H-function \cite[A.1]{M-Foxh} to get \eqref{eq:sum_mult_pdf}.
Similarly CDF can be computed as $F_{X_{ }}(x) = \mathcal{L}^{-1} \prod_{i=1}^{N} \frac{M_{Z_i}(s)}{s}$, with which inner integral becomes 
\begin{eqnarray}\label{eq:sum_mult_cdf_inner_int}
&\frac{1}{2\pi \J} \int_{\mathcal{L}} s^{\sum_{i=1}^{N}-(\sum_{k=1}^{M}a_{i,k}s_{i,k})-a_{i}-1} e^{sx} ds \nonumber\\&= \frac{x^{\sum_{i=1}^{N}(\sum_{k=1}^{M}a_{i,k}s_{i,k})+a_{i}}}{\Gamma(\sum_{i=1}^{N}(\sum_{k=1}^{M}a_{i,k}s_{i,k})+a_{i}+1)} 
\end{eqnarray}

We substitute it back and apply the definition of Multi-variate Fox's H-function to get \eqref{eq:sum_mult_cdf}.

\bibliographystyle{IEEEtran}
%\linespread{0.9}
\bibliography{Multi_RISE_full}

\end{document}